\documentstyle[epsf,12pt]{article}
\title{\bf Algebraic  approach to the spectral problem for
the Schr\"odinger equation with  power potentials}
\author{R.N. Faustov, V.O. Galkin \\
{\small \it Russian Academy of Sciences, Scientific Council for
Cybernetics} \\
{\small \it Vavilov Street 40, Moscow 117333, Russia}\\
\smallskip\\
A.V. Tatarintsev, A.S. Vshivtsev\\
{\small \it Moscow State Institute for Radiotechnics, Electronics and
Automatics}\\
{\small \it Prosp. Vernadskogo 78, Moscow 117454, Russia}}

\date{ }
\begin{document}
\maketitle

\begin{abstract}

The method reducing the solution of the Schr\"odinger equation for
several types of power potentials to the solution of the eigenvalue
problem for the infinite system of algebraic equations is developed.
The finite truncation of this system provides  high accuracy results
for low-lying levels. The proposed approach is appropriate both for
analytic calculations and for numerical computations. This method
allows also to determine the spectrum of the Schr\"odinger-like
relativistic equations. The heavy quarkonium (charmonium and
bottomonium) mass spectra for the Cornell potential and the sum of the
Coulomb and oscillator potentials are calculated. The results are in
good agreement with experimental data.
\end{abstract}

\section{Introduction}

The solution of the spectral problem for the Schr\"odinger equation
with spherically symmetrical potentials is an important problem for
the atomic and hadronic spectroscopy. It is well known that
potential models give
a good description of the heavy quarkonium ($\psi$, $\Upsilon$) mass
spectrum \cite{qm}. The interaction potential in such a system is
usually assumed to be of the confining type. The example is
the Cornell potential which contains two terms, one is responsible for
the Coulomb interaction dominating at small distances and the other
corresponds to the string-like interaction leading to the confinement.

The problem of calculating the energy spectrum of the Schr\"odinger
equation with different potentials is attracting the attention of
physicists  for a long time and is being solved by many methods: by
the direct numerical solution with fixed boundary conditions for wave
functions, by the variational method and by different modifications
combining analytical and numerical approaches.  The quasiclassical
method  works rather well for a wide class of quantum mechanical
problems \cite{mpsy,mp}. However, most of these approaches does not
allow to  determine explicitly the dependence of the solution  on the
parameters of the potential.

In this paper we  develop further the method proposed in
Refs.~\cite{vns,vs} which accounts for the asymptotic behaviour of the
wave functions at large distances (or small momenta). This approach,
in fact, generalizes the method of integral transformations for the
kernels of special type, determining the correct asymptotic behaviour
of the wave functions.  As it was shown in Refs.~\cite{vns,vs} such a
method is rather efficient for finding the low-lying spectral
levels. These low-lying levels are of a special interest for the meson
and baryon spectroscopy.

Here we give the exact algorithm for the reduction of the radial
Schr\"odinger equation to the algebraic eigenvalue problem, from which
it is possible to obtain the spectrum of the initial equation using
successive approximation method.  This approach is rather simple and
suitable both for analytical calculations and numerical computations
and allows to get approximate analytical expressions for the low-lying
levels. These expressions are suitable for the qualitative analysis of
the dependence of the spectrum on the parameters of the model
potential. It is necessary to note that the resulting approximation
scheme is rather rapidly convergent. This result is of the independent
interest and shows  the reliability of developed approach.

In this paper we consider spherically symmetrical confining potentials
of the following form
\begin{equation}
\label{pot}
U_k(r)=a^2r^k-\frac Zr,\qquad a >0, \quad   k=1,2,
\end{equation}
which are important for hadron physics.

The general construction of the fast converging approximation scheme
with the account of the algebraic symmetry is given in  \cite{vns}.
Here we apply this method for the potentials of the type (\ref{pot}).

The radial part of the Schr\"odinger equation can be presented as
follows:
\begin{equation}
\label{shr}
\left[ -\frac{d^2}{dr^2}+U_{eff}\left( r\right) \right] R\left(
r\right)=\lambda R\left( r\right).
\end{equation}
We assume that the total wave function of the solution contains a
standard angular part as well:
\begin{equation}
\label{wf}
\psi _{nlm}\left( {\bf r}\right) =\frac{R_n\left(
r\right)}rY_{lm}\left( \theta ,\varphi \right).
\end{equation}
In this case the effective potential in (\ref{shr}) will be of a
well-known type:
\begin{equation}
\label{eff}
U_{eff}\left( r\right) = \frac{l\left( l+1\right)}{r^2}+U_k\left(
r\right).
\end{equation}

If the potential $U_k\left( r\right) $ does not contain terms more
singular at the origin than $r^{-1  }$ (as in the case of
(\ref{pot})), the behaviour of the radial part of the wave function
$R\left( r\right) \sim r^{l+1}$ at the origin will be determined by
the first term in (\ref{eff}). Let us make the substitution
$R\left( r\right) \sim r^{l+1} F\left( r\right) $, where   $F\left(
0\right) \neq 0$ . Then the equation can be rewritten in the form
convenient for further transformations
\begin{equation}
\label{eq1}
F^{\prime \prime }+\frac 2r\left( l+1\right) F^{\prime} +\left[
\lambda-U\left( r\right) \right] F=0.
\end{equation}

\section{The spectral problem for potential $U_1(r)$}

In the case of the potential (\ref{pot}) linearly depending on $r$
the constant  $a^2>0$ can be removed by the simultaneous change
of the variable in eq.~(\ref{eq1})  $r\rightarrow a^{-2/3}r$ and
introduction of the dimentionless combinations of parameters
$\lambda ^{*}=\lambda a^{-4/3}$,   $z^{*}=za^{-2/3}$. The solutions of
eq.~(\ref{eq1}) are sought in the form
\begin{equation}
\label{raw1}
F\left( r\right)
=\sum^{\infty}_{k=0}\frac{a_k}{f_k}r^k,\qquad
f_k=\left( \frac 23\right) ^{2k/3}\frac{\Gamma
\left(\frac{2k+4l+5}3\right) }{\Gamma \left( \frac{4l+5}3\right)} .
\end{equation}

Coefficients  $f_k$ are chosen from  the considerations of
symmetrization of the Jacobi matrix, constructed from the coefficients
$a_k$, and  receiving of the correct asymptotical behaviour of the
wave function ad infinitum  $R\left( r\right) \sim \exp \left[ -\frac
23r^{3/2}\right] $, coinciding in this case with the asymptotics of
the Airy function. Moreover, such a choice of coefficients allows to
adjust the kernel of generalized integral Laplace transformation
(Mittag-Leffler transformation). The recursion relation obtained by
substitution of (\ref{raw1}) in (\ref{eq1}) is the following:
\begin{eqnarray}
\label{req1}
& &\left( k+2\right) a_{k+2}-\left(
k+2l+\frac 23\right) a_{k-1}+\left( \frac23\right)
^{1/3}\frac{z^{*}}{\Gamma \left( \frac 13\right)}\,{\rm
B}\!\left(\frac{2k+4l+6}3;\frac 13\right) a_{k+1}\cr
& &+\left( \frac 23\right) ^{-1/3}\frac{\lambda ^{*}}{\Gamma \left(
-\frac13\right) }\,{\rm B}\!\left( \frac{2k+4l+6}3;-\frac 13\right)
a_k=0.
\end{eqnarray}

Introducing generating function   $\Phi \left( w\right)
=\sum^{\infty }_{k=0}a_kw^{2k/3}$  this relation can be transformed to
the integro-differential equation:
\begin{equation}
\label{eq2}
\left[ \left( 1-w^2\right) \Phi ^{\prime }-\frac{4l+5}3w\Phi
\right]+\frac{z^{*}}{\Gamma \left( \frac 13\right) }\left( \frac
23\right)^{4/3}w^{-\frac 13}I+\frac{\lambda^{*}}{\Gamma \left( -\frac
13\right)}\left( \frac23\right) ^{2/3}w^{\frac 13}I^{*}=0,
\end{equation}
where
$$I\left( w\right) =\int^{1}_{0}dt\,
t^{\frac{4l+1}3}\left( 1-t\right) ^{-\frac 23}\Phi \left(wt\right),
$$

$$I^{*}\left( w\right)=\int^{1}_{0}dt\,
t^{\frac{4l+3}3}\left( 1-t\right) ^{-\frac 43}\Phi \left(wt\right). $$

Factoring out explicitly a pole  singularity of the function at
$w=1$,
$$\Phi \left( w\right) =\left( 1-w^2\right)^{\frac{-4l-5}6}H\left(
w\right), $$
and performing the conformal mapping  of the $w$  plane in the unit
circle   $x=\frac{1-w}{1+w}$ for new unknown funcion
$\overline{\Lambda }\left( x\right) $, where $H\left( w\right)\equiv
x^{\frac{4l+5}6}\left( 1-x\right)^{\frac{-4l+2}3}\overline{\Lambda
}\left( x\right) $, we get an equation:
\begin{equation}
\label{eq3}
x\left( 1-x\right) \overline{\Lambda }^{\prime }+\left[
\frac{4l+5}6+\frac{4l-9}6x\right] \overline{\Lambda
}=\frac{2^{2/3}z^{*}}{3^{4/3}\Gamma \left( \frac 13\right)
}\left( 1-x\right)^{-2/3}I\left( x\right) +\frac{\lambda
^{*}}{2^{2/3}3^{2/3}\Gamma \left( -\frac13\right) }\,I^{*}
\left( x\right).
\end{equation}
Here the integrals $I$, $I^{*}$ are connected to the function
$\overline{\Lambda }\left( x\right) $ by the transformations:

$$I\left( x\right) =\int^{1}_{x}\left(
1-u\right) u^{-\frac 23}\left( 1-\frac xu\right)
^{-\frac23}\overline{\Lambda }\left( u\right) du, $$

$$I^{*}\left( x\right)=\int^{1}_{x}\left(
1-u\right)^{\frac 53}u^{-\frac 43}\left( 1-\frac xu\right) ^{-\frac
43}\overline{\Lambda}\left( u\right) du.$$

According to general theory \cite{vns} the necessary asymptotic
behaviour of the wave function is satisfied if $x=0$ is a regular
point of the function  $\overline{\Lambda }\left(
x\right) $. This allows to search for such solutions in the form of
the nonnegative power series:
\begin{equation}
\label{raw2} \overline{\Lambda }\left( x\right)
=\sum^{\infty}_{k=0}A_kx^k.
\end{equation}
Performing the integrals  $I$, $I^{*}$ with the help of the direct and
inverse integral Mellin transformation, it is easy to find the
recursion relation for the coefficients  $A_k$:

\begin{equation}
\label{eq4}
\left( n+\frac 23l+\frac 56\right) A_n-\left( n-\frac23l+\frac 12\right)
A_{n-1}=\overline{z}\sum^{\infty
}_{k=0}F_{nk}A_k+\overline{\lambda
}\sum^{\infty}_{k=0}G_{nk}A_k.
\end{equation}
Here
$$F_{nk}=\frac 1{n!}\sum^{n}_{m=0}C_n^m\left[
\frac{\Gamma \left( n-m+\frac 23\right) }{\Gamma \left( \frac
23\right)}\right] \left[ \frac{\Gamma \left( m+\frac 23\right) }{\Gamma
\left( \frac23\right) }\right] \left[ \frac{\Gamma \left( k-m+\frac
13\right) }{\Gamma\left( \frac 13\right) }\right] \left[ \frac{\Gamma
\left( \frac73\right)}{\Gamma \left( k-m+\frac 73\right) }\right], $$

$$G_{nk}=\frac 1{n!}\left[ \frac{\Gamma \left( \frac 73\right)
}{\Gamma \left(k-n+\frac 73\right) }\right] \left[ \frac{\Gamma \left(
n+\frac43\right)}{\Gamma \left( \frac 43\right) }\right] \left[
\frac{\Gamma\left( k-n-\frac13\right) }{\Gamma \left( -\frac 13\right)
}\right], $$
are semi-infinite matrices and  $\overline{\lambda }=5\cdot
2^{-5/3}3^{-2/3}a^{-4/3}\lambda \Gamma \left( \frac23\right) /\Gamma
\left( \frac 13\right) $,
$\overline{z}=2^{-4/3}3^{2/3}a^{-2/3}z/\Gamma \left( \frac13\right) $
are parameters. The further procedure of getting approximate energy
eigenvalues  $\lambda _n$ consists in the finite truncation of the
matrices $\hat F$, $\hat G$ at the dimensions
$\left( N+1\right) \times \left( N+1\right) $, where $N=0,1,2,\dots$.
The resulting $N+1$ eigenvalues  $\lambda_n$,
$n=0,1,...,N$ correspond to the position of the first levels of the
spectral problem. The higher is the value of  $N$, the lower is the
dispersion value. The accuracy of the lowest levels is the highest. In
Appendix we illustrate the convergence of the proposed
approximation by the exactly solvable examples, which are the limits
of the potential (\ref{pot}). There it is shown that
the proposed approach exactly reproduces the solutions of the
Schr\"odinger equation for the oscilator and the Coulomb potentials.
Our perturbation results for $S$ wave linear potential rapidly
converge to the Airy function zeros. For the ground state the
1\% accuracy is achieved already for $N=1$.

For the nonzero  $z$ values it is necessary to take higher
appproximations in $N$. However, the dependence of the solution on the
potential parameters $a$ and $ z$ is qualitatively reproduced even
for  $N=0,1,2$. For example, for $N=0$ the ground state in this
potential is
\begin{equation}
\label{fo}
\lambda _{0,l}^{\left(
N=0\right) }=1.742a^{4/3}\left( l+\frac 54\right)-2.415a^{2/3}z
\end{equation}

For   $N=1$ we get two series of levels corresponding to the ground
and first exited radial states
\begin{equation}
\label{so}
\bar \lambda _{0,l}^{\left( N=1\right) }=q-\sqrt{\frac D4},\qquad
\bar \lambda_{1,l}^{\left( N=1\right) }=q+\sqrt{\frac D4},
\end{equation}
where
$$q=\frac{13}{18}l-\frac{23}{24}\overline{z}+\frac{43}{36},$$
and
$$\frac
D4=q^2-\left(\frac{31}{36}l
+\frac{113}{144}-\frac{13}{12}\overline{z}\right)\left(
\frac7{12}l+\frac{77}{48}-\frac{5}{6} \overline{z}\right) +\left(
\frac1{12}l+\frac{11}{48}-\frac{11}{42}\overline{z}\right) \left(
\frac{49}{36}l-\frac{49}{144}-\frac{7}{12}\overline{z}\right). $$

For $N=2$ the solutions of cubic equation correspond to three radial
quantum numbers 0,1,2.

It is necessary to point out that the reasonable accuracy of these
analytical expressions can be achieved only in the limited region of
parameters $z<z_{\max }\left( N\right) $,
$a<a_{\max }\left( N\right) $. The boundary values of parameters are
higher for higher values of the matrix truncation parameter $N$. The
estimates of the boundary parameter values can be obtained during
numerical computations. The convergence of this method is guaranteed
by the compact nature of the initial differential operator and its
speed is rather high for carrying numerical computations which do
not need the extensive computer resources.

\section{The spectral problem for potential $U_2\left( r\right) $}

As in the previous case, for the potential $U_2\left(
r\right) $  the substitution  $r\rightarrow \frac r{\sqrt{a}}$,
$z^{*}=\frac Z{\sqrt{a}}$, $\lambda^{*}=\frac \lambda a$ removes the
coefficient of the highest power of $r$. Carrying out the similar
consideration it is easy to get an equation corresponding to
(\ref{eq3}) in the form:
\begin{equation} \label{eq5}
x\left( 1-x\right) \overline{\Lambda }^{\prime}+\left[ \left( \frac
l2+\frac 34-\frac \lambda 4\right) +\left( \frac l2-\frac
34-\frac\lambda 4\right) x\right] \overline{\Lambda }=\frac
12\overline{z}I\left(x\right),
\end{equation}

$$I\left( x\right) =\int^{1}_{x}u^{-\frac
12}\left( 1-\frac xu\right) ^{-\frac 13}\overline{\Lambda }\left(
u\right)du,\qquad   \overline{z}=\frac{Z^{*}}{\sqrt{2}\Gamma \left(
\frac12\right) }.$$
The recursion relation for the coefficients of the decomposition
$\overline{\Lambda }\left( x\right)
=\sum^{\infty}_{k=0}A_kx^k$ is
\begin{equation}
\label{eq6}
\left( n+\frac l2+\frac 34-\frac \lambda 4\right) A_n-\left( n-\frac
l2-\frac 14-\frac \lambda 4\right)
A_{n1}=\overline{z}\sum^{\infty}_{k=0}F_{nk}A_k,
\end{equation}
$$F_{nk}=\frac 1{2n!}\left[ \frac{\Gamma \left( n+\frac
12\right) }{\Gamma\left( \frac 12\right) }\right] \left[ \frac 1{k-n+\frac
12}\right]. $$

The zero order approximation gives:
\begin{equation}
\label{oscil}
\lambda _{0,l}^{\left( N=0\right) }=a\left( 2l+3\right)
-\sqrt{\frac{8a}\pi}z.
\end{equation}
Note that although this approximation for $Z=0$ gives the exact
result, its accuracy with the growth of $Z$  is decreased. Thus for
nonzero $Z$ it is necessary to take higher approximations.

The limiting exactly solvable cases $Z\to 0$, $a\to 0$ are considered
in Appendix.

\section{Heavy quarkonia mass spectrum }

In this section we apply the developed method of the solution of the
Schr\"odinger equation to the calculation of the mass  spectra of
heavy quarkonia -- mesons consisting from heavy quark and antiquark
($\Upsilon(b\bar b)$, $\psi(c\bar c)$). The main assumption of the
nonrelativistic quark model \cite{qm}, which is widely used for the
heavy hadron description, is the application of the
nonrelativistic approximation. This approximation works better with
the increase of the constituent quark mass. The calculations in the
framework of relativistic quark models \cite{rel} indicate that
nonrelativistic approximation gives a good description of the static
properties of heavy quarkonia (such as mass spectrum, radius etc.),
while for dynamical properties (such as decays) the account of
relativistic corrections is considerably more important. In the
nonrelativistic quark model quarkonium is described by the
Schr\"odinger equation with local spherically symmetrical potential,
which  ensures quark confinement:
\begin{equation}
\label{schrod}
-\frac{1}{2\mu}\Delta\Psi({\bf r}) +[V({\bf r})-
E]\Psi({\bf r})=0,
\end{equation}
where $\mu=m_1m_2/(m_1+m_2)$ is the reduced mass for the bound system
consisting from the quark and atiquark with masses  $m_1$ and $m_2$.

One of the most popular is the Cornell potential \cite{cornell} which
accounts for asymptotic freedom in QCD at small distances and the
linear rise of the potential with the increasing distance between
quark and antiquark:
\begin{equation}
\label{corn}
V(r)=-\frac 43 \frac{\alpha_s}{r} + Ar+B,
\end{equation}
where $\alpha_s$ is a strong interaction constant.
Sometimes the quadratic confinement is used:
\begin{equation}
\label{osc}
V(r)=-\frac 43\frac{\alpha_s}{r} +Ar^2+B.
\end{equation}
The equation (\ref{schrod}) with potentials (\ref{corn}), (\ref{osc})
reduces to the radial equation (\ref{shr})  with the potential
(\ref{eff}) after substitution:
$$ \lambda=2\mu(E-B), \quad a^2=2\mu A, \quad Z=\frac 83\alpha_s\mu.$$

According to our method, the solution of (\ref{schrod})
is reduced to the solution of the systems of algebraic equations
(\ref{eq4}) for the Cornell potential (\ref{corn}) and
(\ref{eq6}) for potential (\ref{osc}). Bottomonium ($\Upsilon(b\bar
b)$) and charmonium ($\Psi(c\bar c)$) mass spectra, obtained by the
final truncation of the matrices in (\ref{eq4}) and (\ref{eq6}) at the
size $9\times 9$ ($N=8$)  are given in comparison with experimental
data \cite{pdg} in Tables~\ref{schbot},~\ref{schchar}. We use the
standard notations for the level centers of gravity: $(n+1)L$, where
$n$ is the radial quantum number.

\begin{table}[hbt]
\caption{Bottomonium ($\Upsilon(b\bar b)$) mass spectrum calculated
for the sum of linear confining potential with the Coulomb potential
and for the sum of oscillator potential with the Coulomb potential
(for $N=8$). The parameters of the Coulomb+linear potential are $
A=0.18$ GeV$^2$, $B=-0.29$ GeV, $\alpha_s=0.39$, $m_b=4.93$ GeV. The
parameters of the Coulomb+oscillator potential are
$A=0.174$ GeV$^3$, $B=-0.05$ GeV, $\alpha_s=0.345$, $m_b=4.95$ GeV.}
\label{schbot}

\begin{center}
\begin{tabular}{|c|ccc|}
\hline
State &Coulomb+linear &Coulomb+oscillator & Experiment\\
\hline
$1S$& 9.447& 9.447 & 9.4604$^*$\\
$2S$& 10.012 & 10.007 & 10.023$^*$\\
$3S$& 10.353 & 10.389 & 10.355$^*$\\
$4S$& 10.629 & 10.742 & 10.580$^*$\\
$1P$& 9.900 & 9.898 &9.900\\
$2P$& 10.260 & 10.259 & 10.260\\
$3P$& 10.544 & 10.593 &\\
$1D$& 10.155 & 10.147 & \\
$2D$& 10.448 & 10.486 & \\
\hline
\end{tabular}

{\small $^*$ -- ${}^3S_1$ state.}

\end{center}
\end{table}

\begin{table}[hbt]
\caption{Charmonium ($\psi(c\bar c)$) mass spectrum calculated
for the sum of linear confining potential with the Coulomb potential
and for the sum of oscillator potential with the Coulomb potential
(for $N=8$). The parameters of the Coulomb+linear potential are $
A=0.18$ GeV$^2$, $B=-0.29$ GeV, $\alpha_s=0.47$, $m_c=1.56$ GeV. The
parameters of the Coulomb+oscillator potential are
$A=0.174$ GeV$^3$, $B=-0.05$ GeV, $\alpha_s=0.345$, $m_b=1.55$ GeV.}

\label{schchar}

\begin{center}
\begin{tabular}{|c|ccc|}
\hline
State &Coulomb+linear &Coulomb+oscillator & Experiment\\
\hline
$1S$& 3.068& 3.070 & 3.0675\\
$2S$& 3.697 & 3.730 & 3.663\\
$3S$& 4.144 & 4.331 & 4.159$^*$\\
$1P$& 3.526 & 3.508 & 3.525\\
$2P$& 3.993 & 4.095 & \\
$3P$& 4.383 & 4.670 & \\
$1D$& 3.829 & 3.841 & 3.770$^{**}$ \\
$2D$& 4.234 & 4.415 & \\
\hline
\end{tabular}

{\small $^*$ -- ${}^3S_1$ state,
$^{**}$ -- ${}^3D_1$ state.}

\end{center}
\end{table}

We can see from Tables~\ref{schbot},~\ref{schchar} that by fitting the
parameters of potentials (\ref{corn}) and (\ref{osc}) it is possible
to get a good agreement with the low-lying level masses of charmonium
and bottomonium for both potentials. Nevertheless, for higher radial
excitations  the discrepancy becomes more pronounced. Unfortunately,
these excited  levels are near the threshold of the open charm
and bottom  production, which makes the comparison with experimental
data more complicated.

The important advantage of the developed method is the possibility to
explicitly determine the dependence of the spectrum on the parameters
of the interquark interaction potential. On Figs.~1-3 we plot the
dependence of the binding energy (in GeV) on the quark mass $m_Q$ (in
GeV) and the Coulomb potential parameter $\alpha_s$ for the Cornell
potential (\ref{corn}). On Figs.~4-6 we present the similar
dependences for the sum of the Coulomb and oscillator potentials
(\ref{osc}).

\section{Mass spectrum of the relativistic Schr\"odinger-like
equation}

There are several different equations for
relativistic two-particle bound states \cite{bs}-\cite{ks}.  Most of
these relativistic equations are the nonlocal integro-differential
equations. However, some of them can be rationalized and reduced to a
local Schr\"odinger-like equation \cite{by,mf}, which in the center of
mass frame can be written as follows:
\begin{equation}
\label{rel}
\left( \frac{{\bf p}^2 }{2\mu_R(M)}+V(M,r)\right)\Psi({\bf
r})= E(M)\Psi({\bf r}),
\end{equation}
where ${\bf p}^2=-\nabla^2=-\Delta$; $E(M)$ and
$\mu_R(M)$ are some functions of the bound state mass $M$; potential
$V(M,r)$ can also depend on $M$.

Equation (\ref{rel}) differes from the ordinary Schr\"odinger equation
(\ref{schrod}) by the dependence of the reduced mass $\mu_R$ and
potential on  energy eigenvalues, which are nontrivial functions of
the bound state mass. In the case of heavy quarkonia the problem can
be simplified by taking the nonorelativistic limit of  potential in
eq.~(\ref{rel})  as the initial approximation. Solving the obtained
equation one exactly accounts for relativistic kinematics. The
relativistic dynamical effects can be taken into account by
calculating  corrections to the potential using perturbation
theory.

The explicit dependence of the spectrum of the Schr\"odinger equation
(\ref{schrod}) on  parameters obtained here also allows the
calculation in the case when the reduced mass  $\mu_R$  and  energy
eigenvalue $E$ depend in the complicated way on the bound state
mass $M$. Thus it is possible to solve the relativistic equation of
the type (\ref{rel}) with the Cornell potential (\ref{corn}).
Substituting  obtained eigenvalues in the initial
equation it is possible to reconstruct the wave function and to
take into account dynamical relativistic corrections using
perturbation theory.

The quasipotential equation \cite{lt}, which is widely used for the
description of bound states in quantum field theory (for application
to meson properties see \cite{fg}), can be rationalized and
transformed  to the local equation \cite{mf} of the type (\ref{rel}).
This equation has the form:
\begin{equation}
\label{quas}
\left(\frac{b^2(M)}{2\mu_{R}}-\frac{{\bf
p}^2}{2\mu_{R}}\right)\Psi_{M}({\bf p})=\int\frac{d^3 q}{(2\pi)^3}
V({\bf p,q};M)\Psi_{M}({\bf q}),
\end{equation}
where the relativistic reduced mass
\begin{equation}
\label{mu}
\mu_{R}=\frac{M^4-(m^2_1-m^2_2)^2}{4M^3};
\end{equation}
and the square of relative momentum on the mass shell
\begin{equation}
\label{b}
b^2(M)=\frac{[M^2-(m_1+m_2)^2][M^2-(m_1-m_2)^2]}{4M^2},
\end{equation}
$V({\bf p},{\bf q};M)$ is a quasipotential operator.
Converting to the coordinate representation it is easy to see
that quasipotential equation is of the type (\ref{rel}) with
\begin{equation}
\label{em}
E(M)=\frac{b^2(M)}{2\mu_R}=\frac{M[M^2-(m_1+m_2)^2][M^2-(m_1-m_2)^2]}{
2M[M^4-(m_1^2-m_2^2)^2]}.
\end{equation}

The solution of  quasipotential equation (\ref{quas}) with the
Cornell potential can be found substituting relativistic reduced mass
(\ref{mu}) and $E$ from (\ref{em}) in  equation (\ref{eq4}).
Note that for  fixed $l$ the number of   determinant
zeros, corresponding to the bound state masses, coincides with the
matrix order. This is a nontrivial result\footnote{The resulting
equation for bound state masses contains fractional powers.}  showing
that our approach gives the correct number of states even for the
complicated mass dependence of energy and reduced mass.
The calculated bottomonium and charmonium masses are given in
Tables~\ref{quasb},~\ref{quasc} together with the results of numerical
solution of the quasipotential equation with the account of
spin-independent relativistic corrections of order $v^2/c^2$
\cite{mass} and in comparison with experimental data \cite{pdg}.  From
these tables we see that our results for  level centers of gravity,
calculated using quasipotential equation, differ from experimental
data by no more than  $1\%$ for bottomonium and  $3\%$ for charmonium.
The account of spin-independent relativistic corrections of order
$v^2/c^2$ \cite{mass} further improves the agreement with experiment.

\begin{table}[hbt]
\caption{ Bottomonium  ($\Upsilon(b\bar b)$) mass spectrum calculated
with the quasipotential equation for the sum of linear confining
potential with the Coulomb potential (for $N=8$). NR -- without
account of relativistic corrections to the quasipotential. SI -- with
the account of spin-independent relativistic corrections of order $v^2/c^2$
to the quasipotential. Potential parameters are $ A=0.18$ GeV$^2$,
$B=-0.30$ GeV, $\alpha_s=0.26$, $m_b=4.88$ GeV.}
\label{quasb}

\begin{center}
\begin{tabular}{|c|ccc|}
\hline
State& NR & SI \cite{mass} & Experiment \\
\hline
$1S$& 9.520& 9.447 & 9.4604$^*$\\
$2S$& 9.990 & 10.018 & 10.023$^*$\\
$3S$& 10.308 & 10.349 & 10.355$^*$\\
$4S$& 10.575 & 10.590 & 10.580$^*$\\
$1P$& 9.881 & 9.894 &9.900\\
$2P$& 10.211 & 10.259 & 10.260\\
$1D$& 10.102 & 10.157 & \\
$2D$& 10.386 & 10.448 & \\
\hline
\end{tabular}

{\small $^*$ -- ${}^3S_1$ state.}

\end{center}
\end{table}

\begin{table}[hbt]
\caption{ Charmonium  ($\psi(c\bar c)$) mass spectrum calculated
with the quasipotential equation for the sum of linear confining
potential with the Coulomb potential (for $N=8$). NR -- without
account of relativistic corrections to the quasipotential. SI -- with
the account of spin-independent relativistic corrections of order $v^2/c^2$
to the quasipotential. Potential parameters are $ A=0.18$ GeV$^2$,
$B=-0.30$ GeV, $\alpha_s=0.32$, $m_b=1.55$ GeV.}

\label{quasc}

\begin{center}
\begin{tabular}{|c|ccc|}
\hline
State& NR & SI \cite{mass} & Experiment \\
\hline
$1S$& 3.155& 3.065 & 3.0675\\
$2S$& 3.718 & 3.669 & 3.663\\
$1P$& 3.546 & 3.517 & 3.525\\
$1D$& 3.833 & 3.808 & 3.770$^*$ \\
\hline
\end{tabular}

{\small $^*$ -- ${}^3D_1$ state.}

\end{center}
\end{table}

\section{Conclusions}

In this paper the approach reducing the solution of the Schr\"odinger
equation for power potentials to the eigenvalue problem for infinite
system of algebraic equations is proposed. Such potentials are widely
used for heavy quarkonium mass spectrum calculations. The finite
truncation (even for small $N$) of this system provides a high
accuracy values for the low-lying levels. The explicit dependence of
the determinant of the system on potential parameters allows  easy
determination of the solution spectrum dependence on these parameters.
It is necessary to note that by substituting  found eigenvalues in the
initial equation it is possible to reconstruct the wave function,
which has the number of zeros coinciding with the radial excitation
number (this is in agreement with the known theorem on the wave
function zero number).

We have investigated the particular cases of the studied potentials
for which the Schr\"odinger equation has exact solutions. It is shown
that for the Coulomb and oscillator potentials the developed method
reproduces the exact spectrum. For the linear rising potential, in
case $l=0$ ($S$ states), our solutions rapidly converge to the Airy
function zeros.

The mass spectra of charmonium and bottomonium for the Cornell
potential (\ref{corn}) and the sum of the Coulomb and oscillator
potentials (\ref{osc}) are calculated on the basis of the developed
approach.  The dependence of spectra on the quark mass $m_Q$ and
strong interaction  coupling constant $\alpha_s$ is investigated. It
is shown that the proposed method allows to determine mass spectra of
relativistic equations which can be presented in Schr\"odinger-like
form. The mass spectra of charmonium and bottomonium for the Cornell
potential are calculated on the basis of the relativistic
quasipotential equation. The obtained results are in accord with
experimental data.

\medskip
\noindent { \Large \bf Acknowledgments}
\smallskip

\noindent
We are grateful to A.I. Aptekarev, A.M. Badalyan, A.V.
Borisov, V.Ch. Zhukovsky, N.N. Nekchoroshev, A.Yu. Simonov, V.N.
Sorokin and E.T. Shavgulidze for useful discussions. The work of
R.N.F. and V.O.G. was supported in part by Russian Foundation for
Fundamental Research under Grant No.~96-02-17171.

\appendix
\setcounter{equation}{0}
\renewcommand{\theequation}{\thesection.\arabic{equation}}
\section{ The limiting cases of the potential (\ref{pot})
for which the Schr\"odinger equation has exact solutions}
In this appendix we use our method to calculate the spectrum of the
Schr\"odinger equation for the limiting cases $Z\to 0$ or $a\to 0$ of
the potential (\ref{pot}) and compare the obtained results with the
exact ones.

\subsection{The spectral problem for oscillator potentail}
The oscillator potential is the limiting case $Z= 0$ of $U_2(r)$
potential. Setting $\bar z=0$ in (\ref{eq6}) we can find the exact
solution of this equation, which reproduce the spectrum of the
spherically symmetrical oscillator
$$\lambda _n=\left( 4n+2l+3\right)
a=2a\left(N_m+\frac 32\right),$$
where $N_m$ is the main quantum number.

\subsection{The spectral problem for linear potential}
The linear potential is the limiting case $Z= 0$ of $U_1(r)$
potential. Setting also $l=0$ ($S$ states) in
(\ref{eq4}) we get the eigenvalues for the spherical linear
potential  $U_{eff}=r$. As it is well known, these solutions
should coincide with the zeros of the Airy function \cite{as}. In
Table~{\ref{airy}} we present the numerical results for the lowest
eigenvalues of eq.~(\ref{eq4}) with $Z=0$, $l=0$ versus the truncation
index $N$. We see that for  $N$ values greater than the radial quantum
number $n$ by 3 the calculated eigenvalues agree with the exact ones
within 1\% accuracy.  This allows to conclude that the convergence
rate of proposed procedure is rather high.

\begin{table}[hbt]
\caption{The comparison of the $S$ state solutions of eq.~(11) for the
potential $V(r)=r$ with the exact ones determined by the zeros of the
Airy ($Ai$)  function. $n$ is a radial quantum number,  ($N+1$)
is the truncation index of matrices $ \hat F$ and $\hat G$ in
eq.~(11).}
\label{airy}
\begin{center}
\begin{tabular}{|c|cccccc|}
\hline
     & \multicolumn{6}{|c|}{$n$}\\
     \cline{2-7}
$N$  & 0 &1& 2& 3& 4& 5\\
\hline
0 & 2.17747 & & & & & \\
1 & 2.33762 & 3.90446 & & & &\\
2 & 2.32928 & 4.19704 & 5.49815 & & & \\
3 & 2.33298 & 4.06720 & 5.87930 & 7.12835 & & \\
4 & 2.33464 & 4.08297 & 5.46807 & 7.57050 & 8.84809 &  \\
5 & 2.33561 & 4.08355 & 5.52374 & 6.70034 & 9.33590 & 10.66484  \\
6 & 2.33623 & 4.08466 & 5.51523 & 6.81654 & 7.85466 & 11.18931 \\
7 & 2.33664 & 4.08538 & 5.51718 & 6.77696 & 8.03348 &$\hphantom{1}$
8.98943  \\
8 & 2.33693 & 4.08588 & 5.51784 & 6.78365 & 7.92349 &$\hphantom{1}$
9.21939 \\
9 & 2.33714 & 4.08626 & 5.51828 & 6.78389 & 7.94408 &$\hphantom{1}$
8.98263 \\
10& 2.33730 & 4.08652 & 5.51905 & 6.78145 & 7.95138 &$\hphantom{1}$
9.01194 \\
\hline Zeros & & & & & & \\
of $Ai$&2.33810& 4.08795 & 5.52056 & 6.78671 & 7.94413 &
$\hphantom{1}$ 9.02265\\
\hline
\end{tabular}
\end{center}
\end{table}

\subsection{The spectral problem for the Coulomb potential}
The Coulomb potential is a limiting case $a=0$ of potential
(\ref{pot}). It is necessary to note that the Coulomb potential in
Schr\"odinger equation with potential (\ref{pot}) does not give the
leading asymptotics for $r\to \infty$, which is determined by the
confining potential. As a result, we cannot get the solution for the
Coulomb problem  just setting  $a=0$ in eq.~(\ref{eq4}) or
eq.~(\ref{eq6}). Therefore, an additional analysis is necessary. The
effective potential in the  equation for the radial part of the wave
function (\ref{shr}) consists in this case of two terms of different
sign:
\begin{equation}
\label{cul}
U_{eff}=\frac{l(l+1)}{r^2} - \frac Zr.
\end{equation}
The bound states in the potential (\ref{cul}) have a negative
eigenvalues. Taking this into account, we make in eq.~(\ref{eq1})
the  substitution  $r\rightarrow r/\sqrt{\left| \lambda
\right| }$, which makes this equation dimensionless, and introduce new
spectral parameter   $\bar{\lambda }=Z/\sqrt{\left| \lambda \right|}$.
After these transformations we get:
\begin{equation}
\label{eqf}
F''+\frac 2r(l+1)F'+\left(-1+\frac{\bar
\lambda }{r}\right)F=0.
\end{equation}

We seek a solution of eq.~(\ref{eqf}) in the form of a series in the
powers of  $r$:
$$F=\sum^{\infty }_{n=0}\frac{A_n}{(n+2l+1)!}r^n.$$
Substituting this decomposition  in eq.~(\ref{eqf}) we find the
recursion  relationship:

\begin{equation}
\label{eq8}
(n+1)A_{n+1}-(n+2l+1)A_{n-1}=-\bar \lambda \,A_n.
\end{equation}

We have made the asymptotic symmetrization of the Jacobi matrix in
eq.~(\ref{eq8}). Introducing the generating function as  $\Phi
(w)=\sum_{n}A_nw^n$, we get the following equation for this
function:
\begin{equation}
\label{eq9}
\left( 1-w^2\right) \Phi ^{\prime }+\left[ \bar \lambda
-2(l+1)w\right] \Phi=0.
\end{equation}

The exact solution of eq.~(\ref{eq9}) is
$$\Phi (w)=(1-w)^{\frac{\bar \lambda}2-l-1}(1+w)^{-\frac{\bar \lambda
}2-l-1}.$$

As it follows from general theory, the behaviour of the generating
function in the point  $w=1$ for given  $\bar
\lambda $ determines the asymptotics of the initial problem wave
function. Thus only for regular function  $\Phi(w)$ in this point, the
wave functions are decreasing with the increase of  $r$, and hence,
they can be normalized in the semi-infinite region. For this
purpose the exponent should be equal to the integer number
$n$. This condition determines the eigenvalue spectrum:   $\bar\lambda
=2(n+l+1)$. Substituting the explicit expression for the
spectral parameter $\bar \lambda $ and taking into account the
negative sign of the binding energy, it is easy to obtain the
well-known Bohr formula for the energy of a particle in the Coulomb
field:
\begin{equation}
\label{bor}
\lambda _N=-\left( \frac Z{2N_m}\right) ^2,\ N_m=n+l+1.
\end{equation}
Here $N_m$ is the main quantum number.

In this appendix we have shown that our method of the Schr\"odinger
equation solution reproduces the exact results for the Coulomb and
oscilattor potentials. For $S$ states in linear potential our
solutions rapidly converge to the zeros of the Airy function.

\section*{Figure Captions}

\noindent {\bf Fig.\ 1} \ {The binding energy as a function of the
quark mass and the strong coupling constant $\alpha_s$ for the Cornell
potential. The parameters of the potential are $A=0.18$ GeV$^2$,
$B=-0.29$ GeV.  Grey patches correspond to  $S$ states, patches with
grid correspond to $P$ states.  }
\smallskip

\noindent {\bf Fig.\ 2} \ {The binding energy as a
function of the strong coupling constant $\alpha_s$  for fixed quark
mass ($m_Q=4.9$ GeV)  for  the Cornell potential. $S$ states are
ploted by solid lines, $P$ states -- by crosses, $D$ states --  by
diamonds.  }
\smallskip

\noindent {\bf Fig.\ 3} \ {The binding energy as a
function of the quark mass for fixed strong coupling constant
($\alpha_s=0.39$)  for  the Cornell potential. $S$ states are ploted
by solid lines, $P$ states -- by crosses, $D$ states --  by diamonds.
}
\smallskip

\noindent {\bf Fig.\ 4} \ {The binding energy as a
function of the quark mass and the strong coupling constant $\alpha_s$
for the sum of the Coulomb and oscillator potentials. The parameters
of the potential are $A=0.174$ GeV$^2$, $B=-0.05$ GeV.  Grey patches
correspond to  $S$ states, patches with grid correspond to $P$ states.
}
\smallskip

\noindent {\bf Fig.\ 5} \ {The binding energy as a
function of the strong coupling constant $\alpha_s$  for fixed quark
mass ($m_Q=4.9$ GeV)  for  the sum of the Coulomb and oscillator
potentials. $S$ states are ploted by solid lines, $P$ states -- by
crosses, $D$ states --  by diamonds.  }
\smallskip

\noindent {\bf Fig.\ 6} \ {The binding energy as a function of the
quark mass for fixed strong coupling constant ($\alpha_s=0.39$)  for
the sum of the Coulomb and oscillator potentials. $S$ states are
ploted by solid lines, $P$ states -- by crosses, $D$ states --  by
diamonds.  }

\epsfysize=625pt \epsfbox{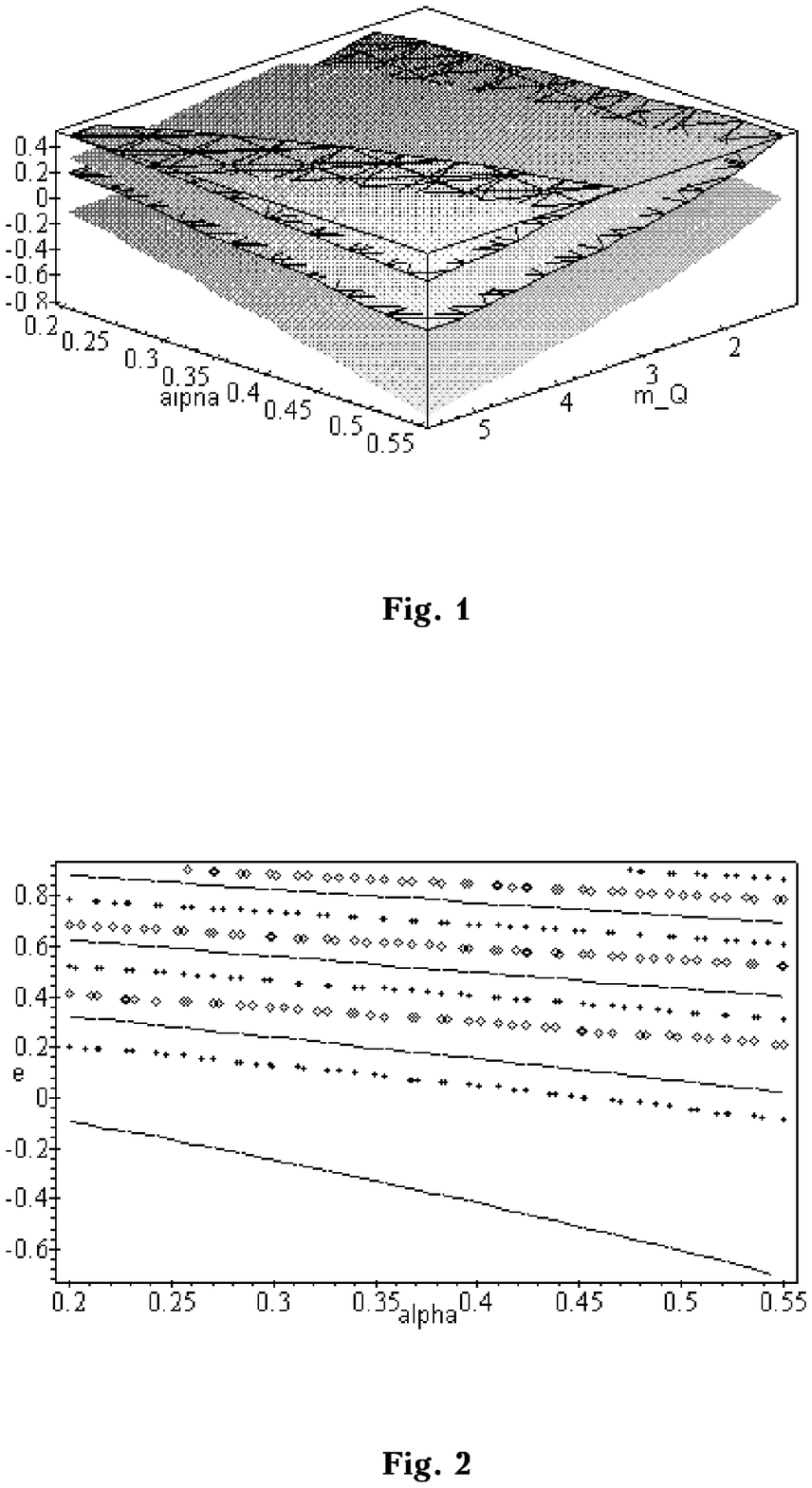}

\epsfysize=625pt \epsfbox{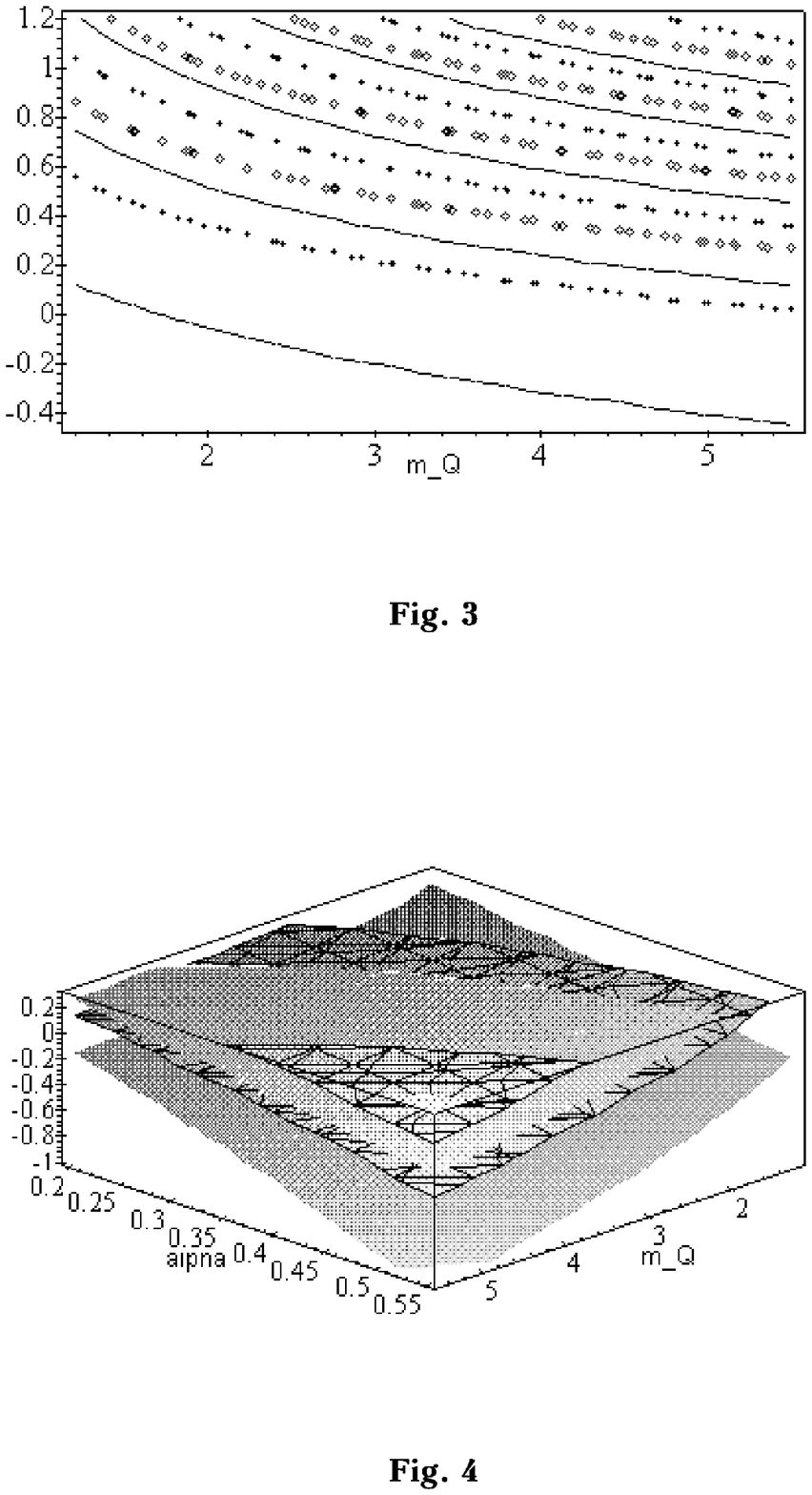}

\epsfysize=625pt \epsfbox{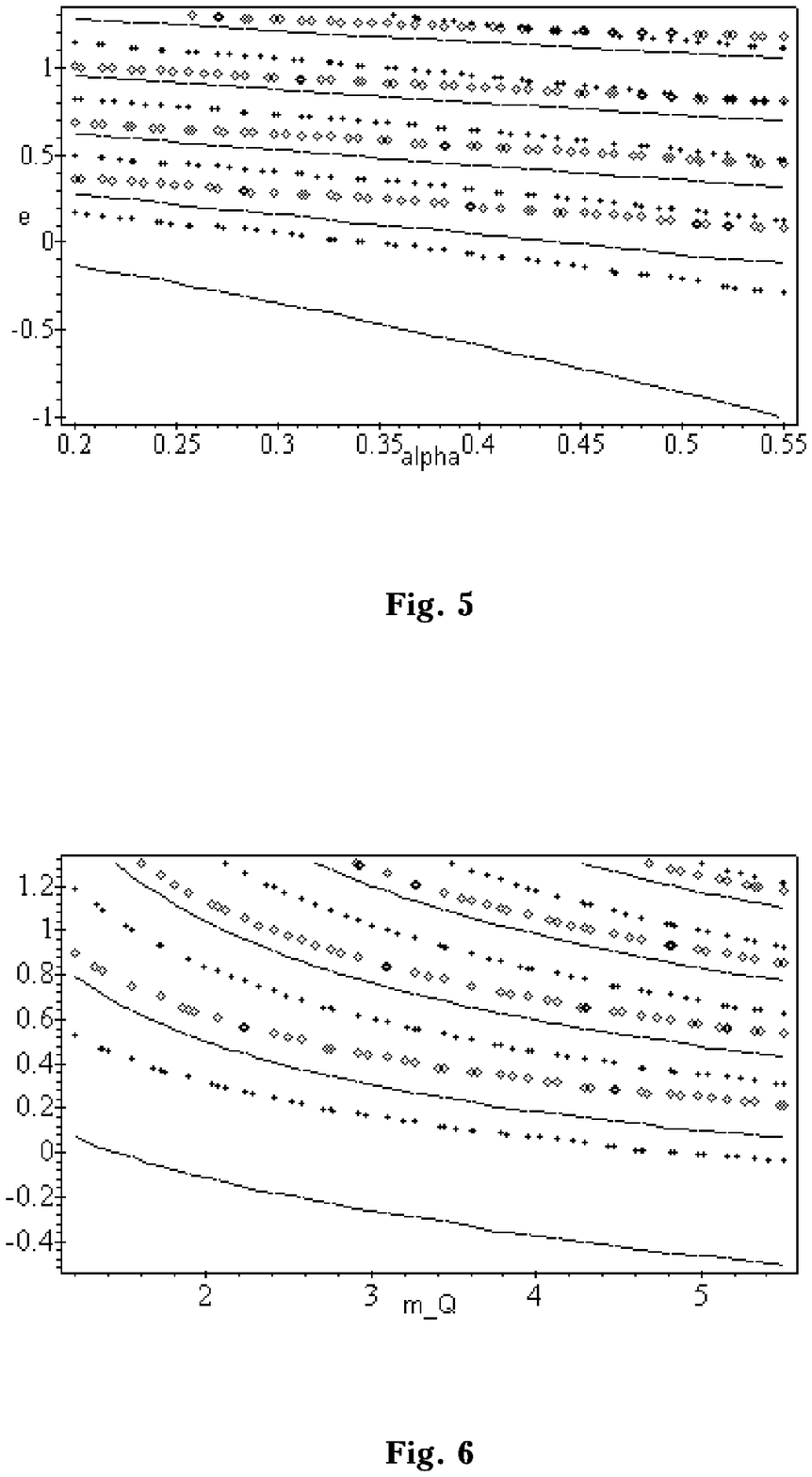}

\end{document}